\documentclass[12pt]{iopart}
\usepackage{amstext}
\usepackage{amsfonts}
\usepackage{amssymb}
\usepackage{amsbsy}
\usepackage{subeqnarray}
\usepackage{setstack}
\usepackage{graphicx}
\usepackage{epsfig}
\usepackage{color}

\newcommand{\abs}[1]{|#1|}              

\newcommand{\BEQ}{\begin{equation}}     
\newcommand{\BEA}{\begin{eqnarray}}
\newcommand{\EEQ}{\end{equation}}       
\newcommand{\EEA}{\end{eqnarray}}
\newcommand{\D}{{\rm d}}                
\newcommand{\II}{{\rm i}}               
\newcommand{\wht}[1]{\widehat{#1}}      

\renewcommand{\vec}[1]{\boldsymbol{#1}} 

\newcommand{\appsection}[1]{\setcounter{equation}{0} \section*{Appendix. #1}
\renewcommand{\theequation}{A\arabic{equation}}
              \renewcommand{\thesection}{A} }

\begin{document}

\title{Ageing in the critical contact process: a Monte Carlo study}
\author{Jos\'e J. Ramasco$^a$, Malte Henkel$^b$, Maria Augusta Santos$^a$ and
Constantino A. da Silva Santos$^a$}
\address{$^a$Departamento de F\'{\i}sica and Centro de F\'{\i}sica do Porto,\\ 
Faculdade de Ci\^{e}ncias, Universidade do Porto,\\ 
Rua do Campo Alegre 687, P--4169-007 Porto, Portugal.}
\address{$^b$Laboratoire de Physique des 
Mat\'eriaux\footnote{Laboratoire associ\'e
au CNRS UMR 7556}, Universit\'e Henri Poincar\'e Nancy I, \\ B.P. 239,
F -- 54506 Vand{\oe}uvre-l\`es-Nancy Cedex, France.}
\eads{\mailto{jjramasc@fc.up.pt},\mailto{henkel@lpm.u-nancy.fr},
\mailto{mpsantos@fc.up.pt},\mailto{cssantos@fc.up.pt}}
\begin{abstract}
The long-time dynamics of the critical contact process which is brought 
suddenly out of an uncorrelated initial state undergoes ageing in close analogy
with quenched magnetic systems. In particular, we show through Monte Carlo
simulations in one and two dimensions and through mean-field theory that 
time-translation invariance is broken and that dynamical scaling holds. We find
that the autocorrelation and autoresponse exponents $\lambda_{\Gamma}$ and
$\lambda_R$ are equal but, in contrast to systems
relaxing to equilibrium, the ageing exponents $a$ and $b$ are distinct. A
recent proposal to define a non-equilibrium temperature through the short-time
limit of the fluctuation-dissipation ratio is therefore not applicable. 
\end{abstract}
\submitto{\JPA}
\date{3$^{\rm rd}$ of June 2004}

\section{Introduction}

It is a well-established fact, theoretically known for many years and also 
experimentally observed,  
that the long-time properties of the ageing 
behaviour in glassy systems can be rationally interpreted in terms of a scaling 
picture which allows to make sense of many peculiarities of these systems
which were once thought to be non-reproducible, see \cite{Stru78}. It has since
been understood that similar effects also occur in non-glassy, 
e.g. simple ferromagnetic systems. From a microscopic point of view, the
basic mechanism appears to be the formation of correlated domains of a
time-dependent typical linear size $\ell(t)$. 
For simple ferromagnets and other non-glassy
systems, it is generally admitted that $\ell(t)\sim t^{1/z}$, where $z$ is the
dynamical exponent. Furthermore, if $\phi(t,\vec{r})$ stands for the order
parameter of such a simple ferromagnet, it turned out that these non-equilibrium
relaxation phenomena are most conveniently studied through two-time quantities 
such as the two-time connected autocorrelator $\Gamma(t,s)$ and the 
two-time linear autoresponse function $R(t,s)$ defined by
\BEQ
\Gamma(t,s) = 
\left\langle \Delta \phi(t,\vec{r})\,\Delta\phi(s,\vec{r})\right\rangle 
\;\; , \;\;
R(t,s) = 
\left.\frac{\delta \langle\phi(t,\vec{r})\rangle}{\delta h(s,\vec{r})}
\right|_{h=0}
\EEQ
where $\Delta\phi(t,\vec{r}) = \phi(t,\vec{r}) - \langle\phi(t)\rangle$ are 
the time-dependent fluctuations of the order parameter 
and $h$ is the magnetic
field conjugate to $\phi$. Causality implies that $R(t,s)=0$ for $t<s$.
By definition, a system is said to undergo {\em ageing}, if $\Gamma(t,s)$
or $R(t,s)$ does not merely depend on the time difference $\tau=t-s$, but 
on both the {\em observation time} $t$ and the {\em waiting time} $s$. 
For recent reviews, see  \cite{Bouc00,Godr02,Cugl02,Cris03,Henk04,Cala04}.

From experimental, numerical and analytical studies on ageing glasses and
ferromagnets it has been learned that these systems
may display dynamical scaling in the long-time limit
\cite{Stru78,Bouc00,Godr02,Cugl02,Cris03,Henk04,Cala04}. Specifically,
consider the two-time functions in the ageing regime $t\gg t_{\rm micro}$,
$s\gg t_{\rm micro}$ and $\tau =t-s\gg t_{\rm micro}$, where $t_{\rm micro}$
is some microscopic time. Then one expects the scaling behaviour
\BEQ \label{1:gl:CRskal}
\Gamma(t,s) \sim s^{-b} f_{\Gamma}(t/s) \;\; , \;\;
R(t,s) \sim s^{-1-a} f_{R}(t/s)
\EEQ
where the scaling functions $f_{\Gamma,R}(y)$ have the following asymptotic
behaviour for $y\to\infty$
\BEQ \label{1:gl:fCR}
f_{\Gamma}(y) \sim y^{-\lambda_\Gamma/z} \;\; , \;\;
f_{R}(y) \sim y^{-\lambda_R/z} .
\EEQ
Here $\lambda_\Gamma$ and $\lambda_R$ are called the autocorrelation 
\cite{Fish88,Huse89} and
autoresponse \cite{Pico02} exponents, respectively.

Up to present, most systems studied were such that they relax towards an
equilibrium steady-state. Here we wish to ask what aspects of the ageing
phenomenology remain if time-dependent systems with a non-equilibrium
steady-state are considered. Besides the obvious question concerning an
eventual scaling behaviour of two-time observables, we shall be specifically
interested in the following two points.

\begin{enumerate}
\item For systems with detailed balance, it is convenient 
to measure the distance of a system from 
equilibrium through the {\em fluctuation-dissipation ratio} \cite{Cugl94}
\BEQ \label{1:gl:FDR}
X(t,s) := T R(t,s) \left( \frac{\partial \Gamma(t,s)}{\partial s}\right)^{-1} .
\EEQ
At equilibrium, the fluctuation-dissipation theorem (FDT) states that 
$X(t,s)=1$. Systems with an equilibrium steady-state may also be 
characterized through the 
limit fluctuation-dissipation ratio  
\BEQ \label{Xinfty}
X_{\infty} = \lim_{s\to\infty} \left( \lim_{t\to\infty} X(t,s) \right) .
\EEQ
If the ageing occurs at the critical temperature $T_c$, $X_{\infty}$ should
be an universal number \cite{Godr00a,Godr00b,Godr02} 
and this has been confirmed in a large variety of systems
in one and two space dimensions \cite{Godr00b,Cala02,Henk03d,Sast03,Chat04}, 
whereas the value $X_{\infty}=0$ 
is expected for temperatures $T<T_c$. The order of the 
limits in (\ref{Xinfty}) is important, since 
$\lim_{t\to\infty}\left(\lim_{s\to\infty} X(t,s)\right)=1$ always. 
On the other hand, Sastre et {\it al.} \cite{Sast03} asserted recently that a
genuinely non-equilibrium temperature might be defined through
\BEQ \label{gl:Tdyn}
\frac{1}{T_{\rm dyn}} := \lim_{s\to\infty} \left( \lim_{t-s\to 0} R(t,s)
\left( \frac{\partial \Gamma(t,s)}{\partial s}\right)^{-1} \right) 
\EEQ
and confirmed this through explicit calculation in the $2D$ 
voter model as well as in other connected spin systems 
that do not satisfy the detailed balance condition. 

Does their definition (\ref{gl:Tdyn}) 
extend to more general non-equilibrium systems (in particular those exhibiting
an absorbing phase)? 
\item For ageing simple magnets, it has been proposed that dynamical scaling
might be generalized to the larger {\em local scale-invariance} 
\cite{Henk02}. Such local scale-transformations, 
with dilatation factors depending 
locally on space and on time, may indeed be constructed for
any given value of $z$. From the condition that $R(t,s)$ transforms covariantly 
under the action of local scale transformations it follows \cite{Henk02,Henk01}
\BEQ \label{1:gl:lsi}
R(t,s) = r_0 \left(\frac{t}{s}\right)^{1+a-\lambda_R/z} 
(t-s)^{-1-a} ,
\EEQ
where $r_0$ is a normalization constant. This prediction has been
confirmed in several models, notably the kinetic Ising model with
Glauber dynamics, both in the bulk \cite{Henk01,Henk03b} and close to
a free surface \cite{Plei04}, the kinetic XY model with a non-conserved
order parameter \cite{Abri04,Pico04}, 
for the Hilhorst-van Leeuwen model \cite{Plei04} 
and several variants of the exactly solvable spherical 
model \cite{Henk01,Godr00b,Cann01,Pico04}.\footnote{A technical complication
arises for the $1D$ Glauber-Ising model where the construction of the
Lie-algebra generators of local scale-invariance must be generalized, see
appendix~C in \cite{Pico04}.} Very recently, the autocorrelation
$\Gamma(t,s)$ could be predicted \cite{Henk04a} from local scale-invariance 
for phase-ordering, where $z=2$. This prediction has been confirmed in the
Glauber-Ising, kinetic spherical and critical voter models and also for the
free random walk \cite{Pico04,Henk04a}. For recent reviews, 
see \cite{Henk03c,Henk04}. 

Is there further evidence in favour of local scale-invariance in statistical 
systems without detailed balance ? 
\end{enumerate}

\noindent Probably the simplest kinetic system far from equilibrium and 
without detailed balance is the celebrated
{\em contact process} which has a steady--state transition in the directed 
percolation universality class. We shall therefore use this model in order
to gain insight into the r\^ole of conditions such as detailed balance 
into the phenomenology of ageing behaviour. A complementary paper studies the
same model through the density-matrix renormalization group \cite{Enss04}. 
In section 2, we recall the definition
of the model. In section 3, we discuss the computation
of correlators either for the non-critical system or at the critical 
point. In section 4, we define the response function and estimate it at 
criticality. After that, in
section 5, we discuss the possibility of a generalization of the FDT for this 
model and in particular whether a non-equilibrium temperature may
be defined through eq.~(\ref{gl:Tdyn}). Finally, we conclude in section 6. 
In the appendix, the mean-field theory of ageing in the contact process is 
discussed.

\section{The model}

The contact process (CP) was originally conceived as
a simple model to describe epidemic disease propagation
\cite{Hinr00,Odor04}. Since then, CP has become a paradigm of Directed
Percolation (DP), arguably the most common universality class of nonequilibrium 
phase transitions with an absorbing state. It is precisely the existence of 
an absorbing state that ensures that this model does not satisfy the 
detailed balance condition. This property stimulates our interest
to try and characterize the model's dynamics in a new way. 

We begin by recalling the definition of the model. In the contact
process, the states of the system are described by a discrete variable, 
$n_i(t)$, defined on the sites $i$ of a hypercubic 
lattice. The possible values of $n_i$ are $1$ or $0$ depending on 
whether the site $i$ is occupied or empty. The dynamics is defined as
follows: for each time step, a site $i$ of the 
lattice is randomly selected. If $i$ is occupied, that particle vanishes
with probability $p$. Otherwise, with probability $1-p$, a new particle is 
created on one of the nearest neighbours of $i$ chosen at random 
(if that new site was still empty). When the control parameter $p$ is varied, 
the model exhibits a continuous phase transition from an
active phase, where the mean density $\langle n(t)\rangle$ 
tends to a constant value $\overline{n}$ in the 
stationary state, to an absorbing phase, with zero final density. Separating
these two phases, there is a critical point located at $p_c = 0.2326746(5)$ 
in $1D$ \cite{Jens93,Dickm98}
(the number in brackets gives the uncertainty in the last digit), and at 
$p_c = 0.37753(1)$ in $2D$ \cite{Dickm98}. We 
have arbitrarily fixed the initial condition of our system such that the
particles are randomly distributed throughout the lattice with 
a mean initial density of $n_0 = \langle n(0)\rangle = 0.8$. The linear system 
sizes used are $L=10^4$ in $1D$ and $L=300$ in $2D$, 
respectively; the typical number of disorder realizations considered 
in simulations
goes up to $2 \times 10^4$ in one dimension and $7 \times 10^3$ in $2D$.  

The introduction of an external
field in the model will be necessary to estimate a response 
function. For this model, the external field $h$ corresponds to an 
additional probability of creation or destruction of particles, depending 
on the sign of $h$.

\section{Two-time autocorrelation functions}

A crucial difference between the CP and usual spin models is that starting 
from a zero-density initial state (corresponding to zero magnetization in 
magnets) is not possible. This configuration corresponds to the absorbing 
state of the CP 
and once the system falls there, no further evolution occurs. Therefore, 
the spatially averaged mean density $\langle n(t)\rangle$ is positive
during the time the systems evolves. For example, it decays as a power law,
$\langle n(t) \rangle \sim t^{-\delta}$, at the critical point $p=p_c$. 
This fact implies that the two--time autocorrelation function defined as
\begin{equation}\label{C}
C(t,s) = \left\langle n(t,\vec{r})\, n(s,\vec{r})\right\rangle , 
\end{equation}    
which is widely used in spin systems or glassy models, is no longer equal 
to the connected autocorrelation function or covariance that reads
\begin{equation}\label{G}
\Gamma(t,s) = \left\langle\Delta n(t,\vec{r})\,
\Delta n(s,\vec{r})\right\rangle,
\end{equation} 
where $\Delta n(t,\vec{r}) = n(t,\vec{r}) -\left\langle n(t)\right\rangle$. 
Both functions are related by means of the expression 
\begin{equation}\label{GG2}
\Gamma(t,s) = C(t,s) -
\left\langle n(t)\right\rangle \, \left\langle n(s)\right\rangle .
\end{equation} 
In eq.~(\ref{GG2}), the second term on the right behaves asymptotically as 
\begin{equation}\label{RTW}
\left\langle n(t)\right\rangle \, \left\langle n(s)\right\rangle
\sim \left\{ \begin{array}{ll} 
(t \, s)^{-\delta} & \mbox{\rm ~~;~ if $p=p_c$} \\
\overline{n}^2  & \mbox{\rm ~~;~ if $p<p_c$}
\end{array} \right.
\end{equation} 
Therefore, for $p\leq p_c$ it follows that 
$C(t,s)$ and $\Gamma(t,s)$ do {\em not} share the same asymptotic behaviour.

\begin{figure}
\begin{center}
\epsfysize=50mm
\epsffile{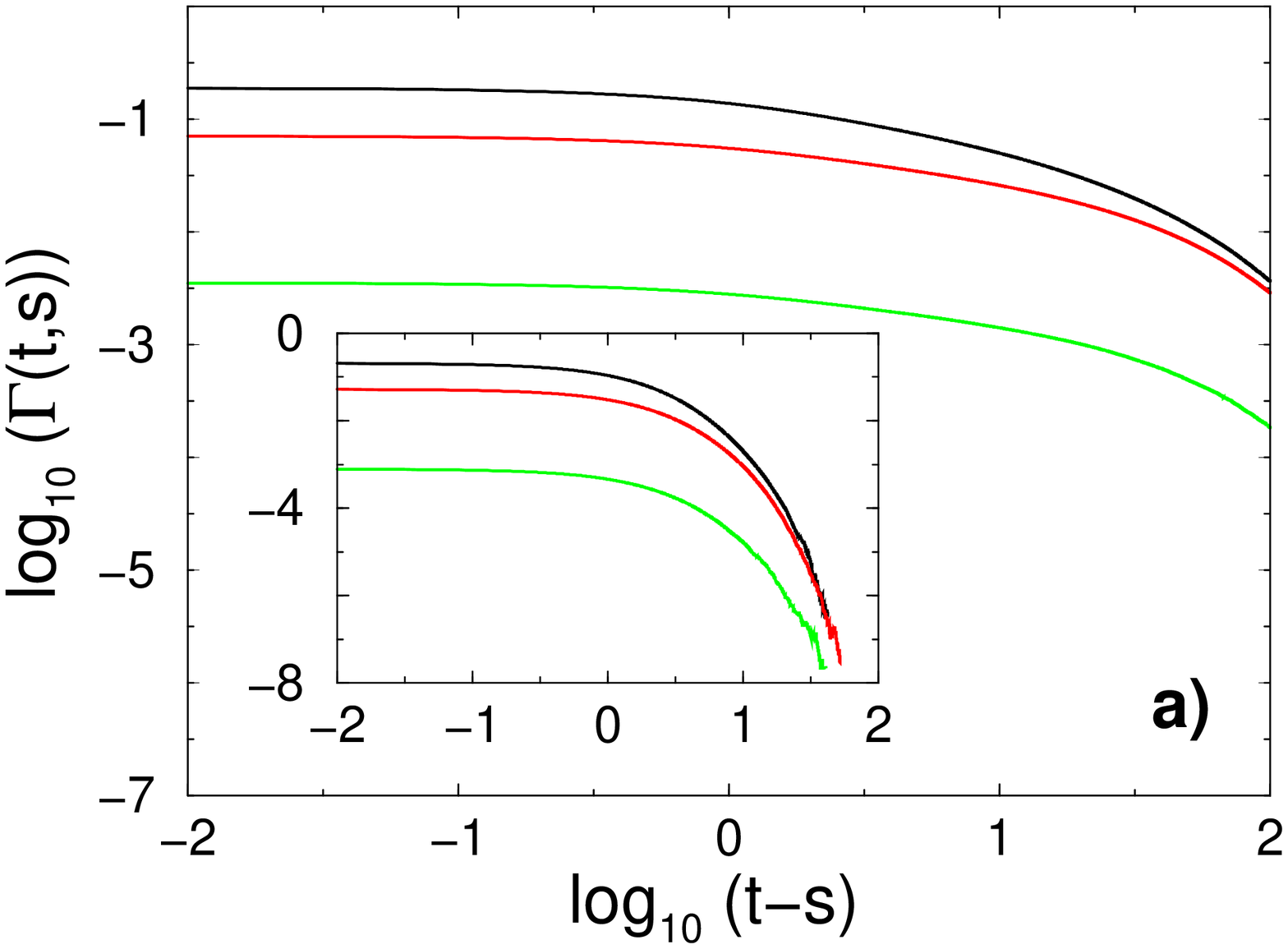}
\epsfysize=50mm
\epsffile{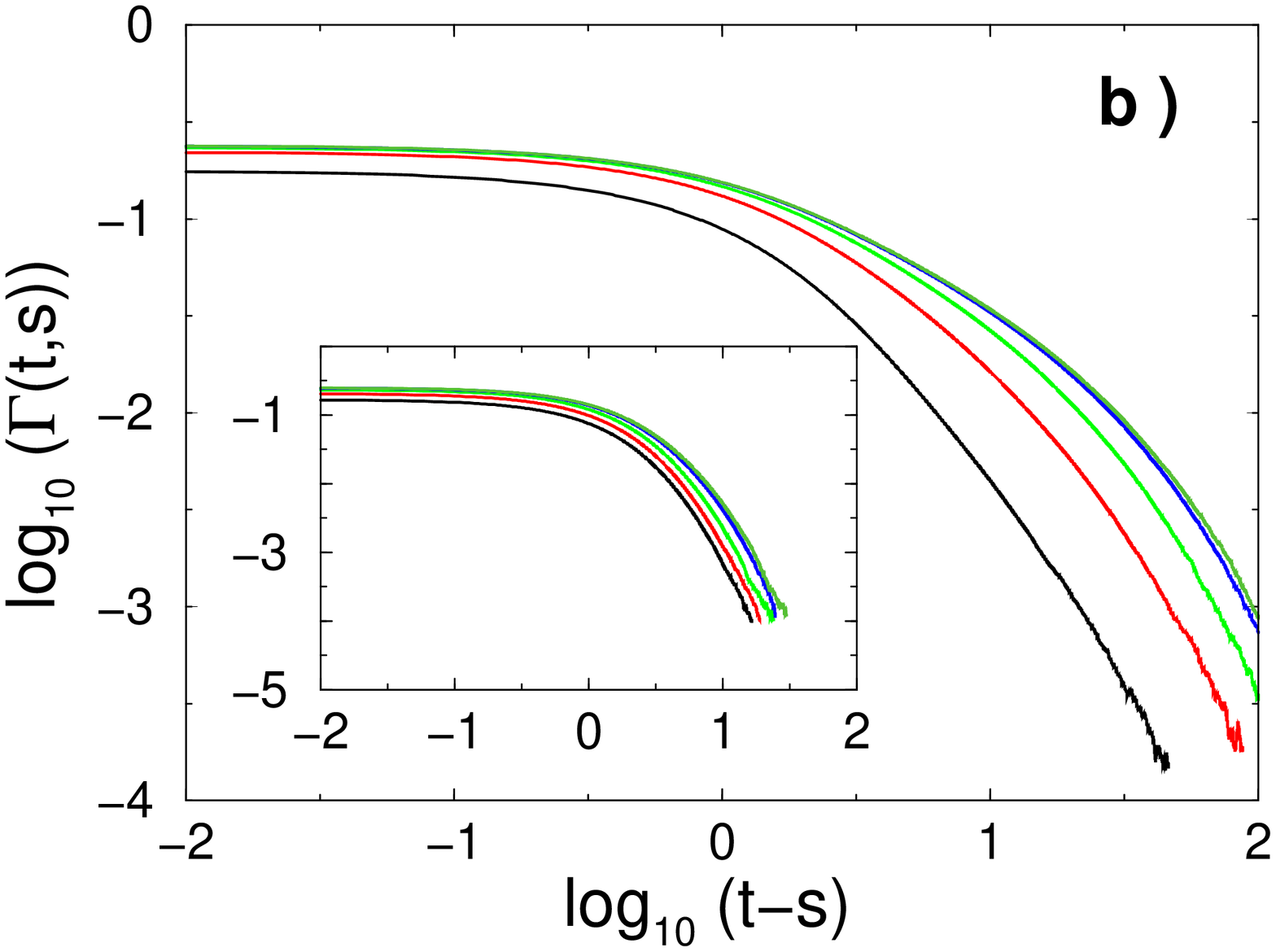}
\caption{Connected autocorrelation function $\Gamma(t,s)$ for the CP for several
values of the waiting time $s$. In a) the steady-state is in the absorbing 
phase. The main plot shows the $1D$ system with $p = 0.286$ and the 
inset shows the $2D$ system with $p = 0.56$. In b) the steady-state is 
in the active phase with $p = 0.2$ in the main plot for $1D$ and $p = 0.286$ 
in the inset for $2D$. The system sizes are $L = 10^4$ in $1D$ and $L = 300$ in
two dimensions. The waiting times in a) are from bottom to top $s=[30,100,300]$
and $s = [3,10,30]$ in the inset. In b), they are, from bottom to top, 
$s =[1,10,30,100,1000]$ and $s = [0.1,1,3,10,100]$ in the inset.}
\label{fig1}
\end{center}
\end{figure} 

Once the correlations have been defined, we are in position to deal with a 
first physical aspect of the temporal evolution of the contact process. 
It is well-known that away from the critical point,
the model has a finite relaxation time towards its steady-state. That is, the 
mean density decays exponentially fast to its stationary value 
$\overline{n}$: zero in the
absorbing phase or a positive quantity in the active one, see \cite{Hinr00}. 
In Figure~\ref{fig1}a, the transient behaviour of $\Gamma(t,s)$ for $t-s$
relatively small is shown for a system evolving to the 
steady-state in the absorbing phase. 
We see that, not only the density, but also its fluctuations with respect to
$\langle n\rangle$ are vanishing, as described by the 
drop of the $\Gamma$--curves. The signature of a finite 
relaxation time may be noted in the persistence of the curve shape 
throughout the system evolution.\footnote{Far above $p_c$, one has for large
$s$ $C(t,s)\sim \Gamma(t,s)\sim \langle n(t)\rangle$, see \cite{Enss04}
for details.} Next, we consider whether a similar behaviour can be
seen in the active phase as well. In Figure~\ref{fig1}b, 
the connected correlation function is displayed for a system in the active 
phase in both $1D$ and $2D$. The curves move upward during a first period of
waiting times before they collapse, which implies that the TTI condition is 
fulfilled in this phase. 

We point out that the observation of TTI at large enough waiting times in
the active phase of the contact process is qualitatively different from what
is found in ferromagnets quenched into their ordered phase. This may be 
understood qualitatively from the Ginzburg-Landau functional 
by observing that for ferromagnets there are two stable competing 
steady-states while for the contact process there is only a single one, 
to which the system relaxes rapidly.

\begin{figure}
\begin{center}
\epsfysize=50mm
\epsffile{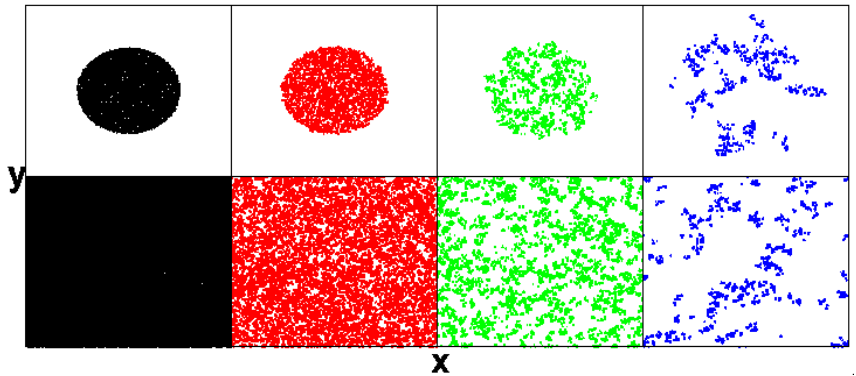}
\caption{Microscopic evolution of clusters in the critical $2D$ contact 
process. The lattice size is, in both series, $1000 \times 1000$. 
The initial condition of the upper plots is a full circle with radius $100$ 
placed in the center of the lattice, while below it is a full lattice. 
The times are $t = [2,20,200,2000]$ for the upper figures 
and $t =[20,200,2000,20000]$ for the bottom snapshots.}
\label{fig2}
\end{center}
\end{figure}

From now on, we concentrate on the behaviour at criticality. In spin systems
with a global symmetry (e.g. an  {\em up-down} symmetry in the Ising model), 
the critical dynamics is based on the growth of spin domains whose average size,
associated to the correlation length, increases as a power law in time,
$\ell(t)\sim t^{1/z}$ where $z$ is the dynamical exponent. The time--evolution 
of typical configurations of the critical CP in two dimensions is shown 
in Figure~\ref{fig2}. 
Note the substantial difference with respect to bulk ferromagnetic spin 
systems: no domain walls are formed and, at the same 
time, the number of occupied sites decreases in time.
This kind of ageing behaviour, not driven by a surface tension, 
has already been described in the $2D$ voter model universality class 
\cite{Dorn01}. Recently, it was pointed out that cluster dilution may
also occur in the early stages of surface ageing of magnetic systems
\cite{Plei04b}. As we shall see, the asymptotic dilution of clusters leads 
to new features in the scaling description of the ageing process of the CP.

\begin{figure} [b!]
\begin{center}
\epsfysize=50mm
\epsffile{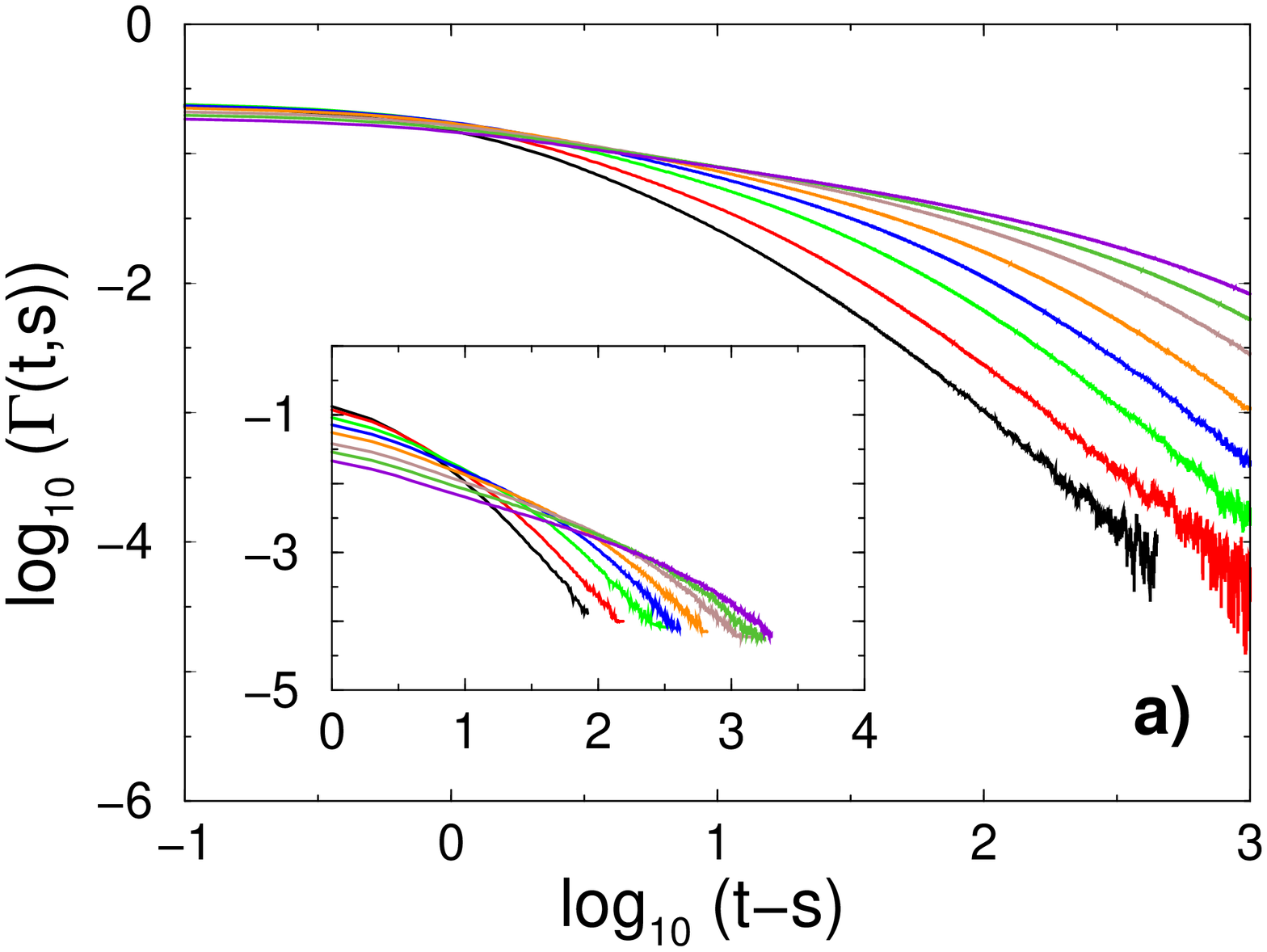}
\epsfysize=50mm
\epsffile{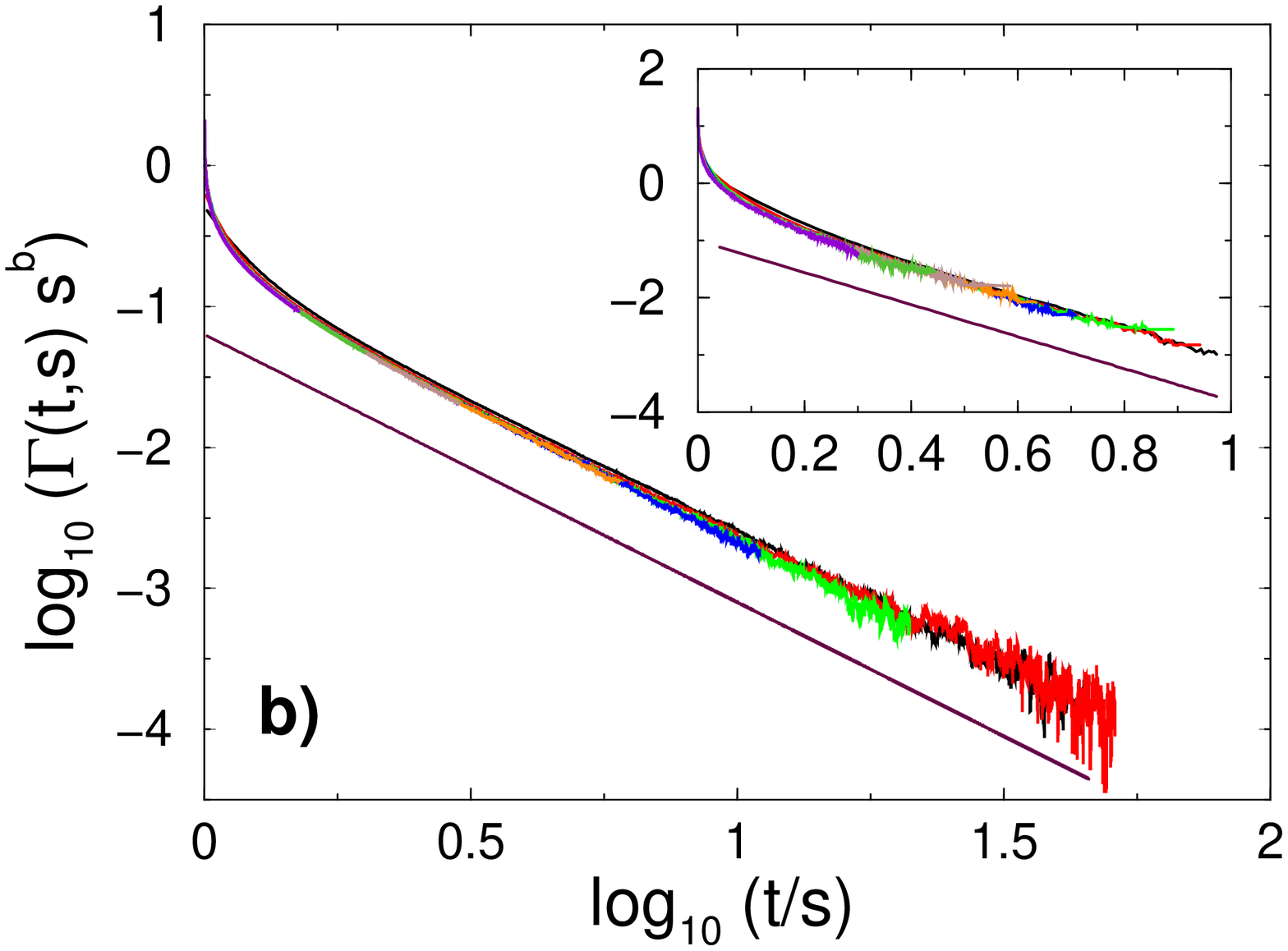}
\caption{a) Evolution of the connected autocorrelation function at criticality. 
The curves in the inset are for two dimensions and the 
main plots for $1D$. The
waiting times are in both cases $s = [10,20,50,100,200,500,1000,2000]$ from 
left to right. In b)
the collapse of the previous curves using Eqs.~(\ref{Gscal}a,\ref{gl:b}) is
shown. The straight lines yield exponent values of $\lambda_\Gamma/z = 1.9$ 
in $1D$ and of $\lambda_\Gamma/z = 2.8$ in $2D$.}
\label{fig3}
\end{center}
\end{figure} 

In analogy with the behaviour of magnetic systems at the critical point, 
a scaling ansatz of the type 
\newpage \typeout{ *** here begins a new page ***}
\begin{subeqnarray}
\label{Gscal}
\Gamma(t,s) &=& s^{-b} f_{\Gamma}(t/s) \;\; \,, \;\; 
f_{\Gamma}(y) \sim y^{-\lambda_{\Gamma}/z} \mbox{\rm ~~if $y\to\infty$} \\
\label{Cscal}
C(t,s) &=& s^{-b} f_C(t/s) \;\; , \;\; 
f_C(y) \sim y^{-\lambda_C/z} \mbox{\rm ~~if $y\to\infty$}
\end{subeqnarray}
may be expected for the correlation functions, where the exponents 
$b,\lambda_C/z$ and $\lambda_{\Gamma}/z$ are to be determined.  
In Figure~\ref{fig3}a, the critical evolution of $\Gamma(t,s)$ is plotted 
for several values of the waiting time $s$, over against $t-s$.  
Time-translation invariance is now clearly broken, and we find ageing 
effects for both one- and two-dimensional systems. 
The data can be collapsed when plotted over against $t/s$, as shown
in Figure~\ref{fig3}b which provides evidence for the validity of the
scaling relation eq.~(\ref{Gscal}a). The values of the 
exponent $b$ used for the collapse 
were obtained from the scaling relation \cite{Wijl98,Muno99}
\begin{equation} \label{gl:b}
b = \frac{2 \, \beta}{\nu_\parallel} =
\frac{(d-2+\eta)}{z} = 2 \, \delta ,
\end{equation}
which remains valid for the DP universality class. Here, $\beta$ is the static
exponent which describes the variation of the order parameter 
 close to criticality $\overline{n}\sim (p_c-p)^{\beta}$, 
$\nu_\parallel$ is the temporal 
correlation-length exponent, $\eta$ is
the static exponent of the order parameter spatial correlations, 
$\delta$ is the exponent controlling the
critical decrease of the mean density  
and $z$ is the dynamic exponent (their values are summarized in table 1).

The exponent $\lambda_\Gamma/z$ may be estimated from the collapse of 
Figure~\ref{fig3}b, with reasonable
numerical accuracy. We find $\lambda_\Gamma/z = 1.9 \pm 0.1$ in $1D$ 
and $\lambda_\Gamma/z = 2.8 \pm 0.3$ in $2D$. If the critical 
CP were a Markov process, these exponents 
might be calculated from the global persistence exponent $\theta_g$ by 
means of the 
scaling relation \cite{Maju96,Hinr98,Muno01} 
\begin{equation}
\frac{\lambda_\Gamma}{z} = \theta_g - \frac{2 (1-d)-\eta}{2 z} ,
\end{equation}
This expression predicts $\lambda_\Gamma/z = 1.98(2)$ in $1D$ and 
$\lambda_\Gamma/z = 3.5(5)$ in $2D$, not too far from our numerical 
measurements. Still, the deviations from the directly measured values
appear to be significant, pointing towards the possible existence of temporal 
long-range correlations and, hence, of an effective non-markovian dynamics. 

\begin{figure}
\begin{center}
\epsfysize=50mm
\epsffile{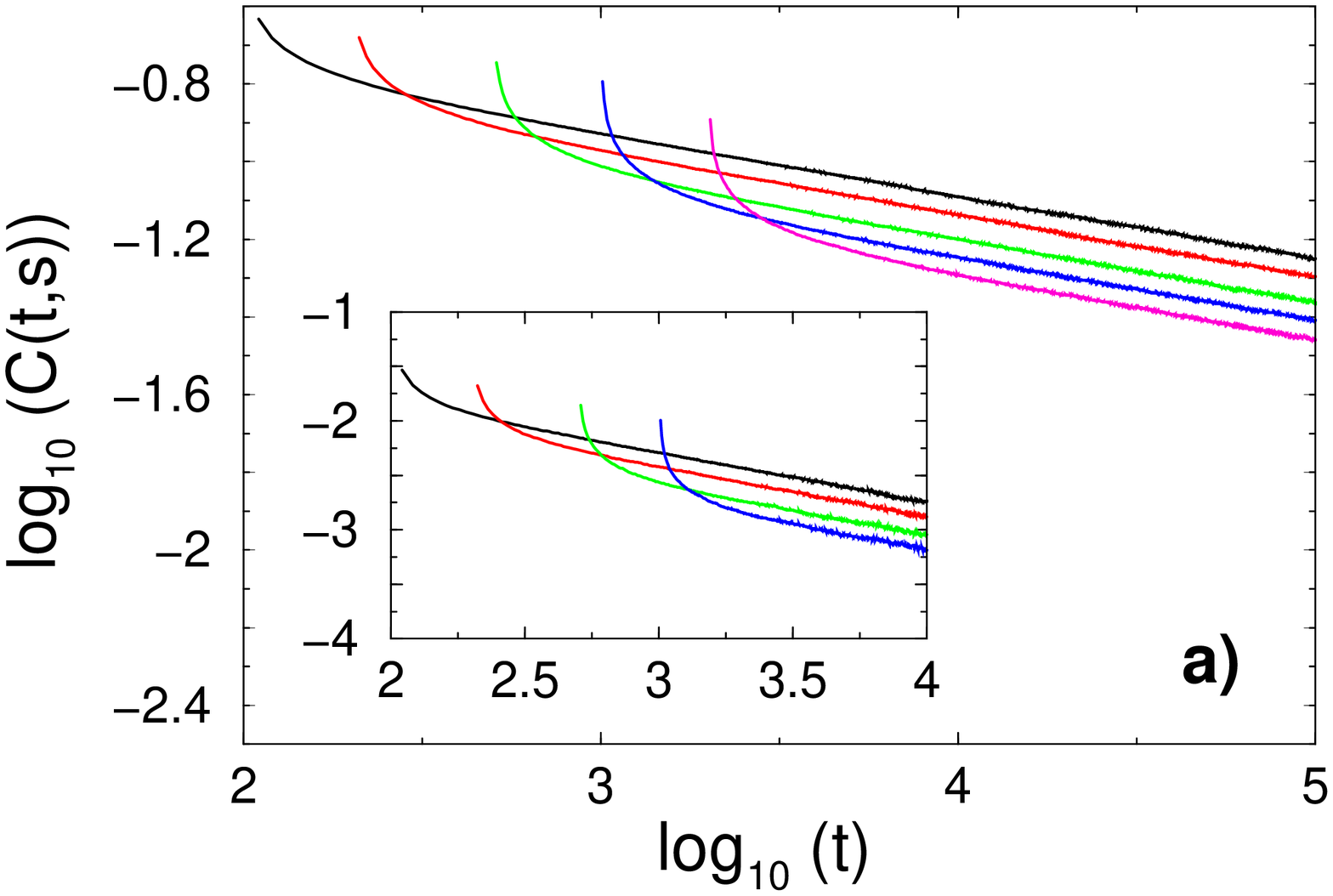}
\epsfysize=50mm
\epsffile{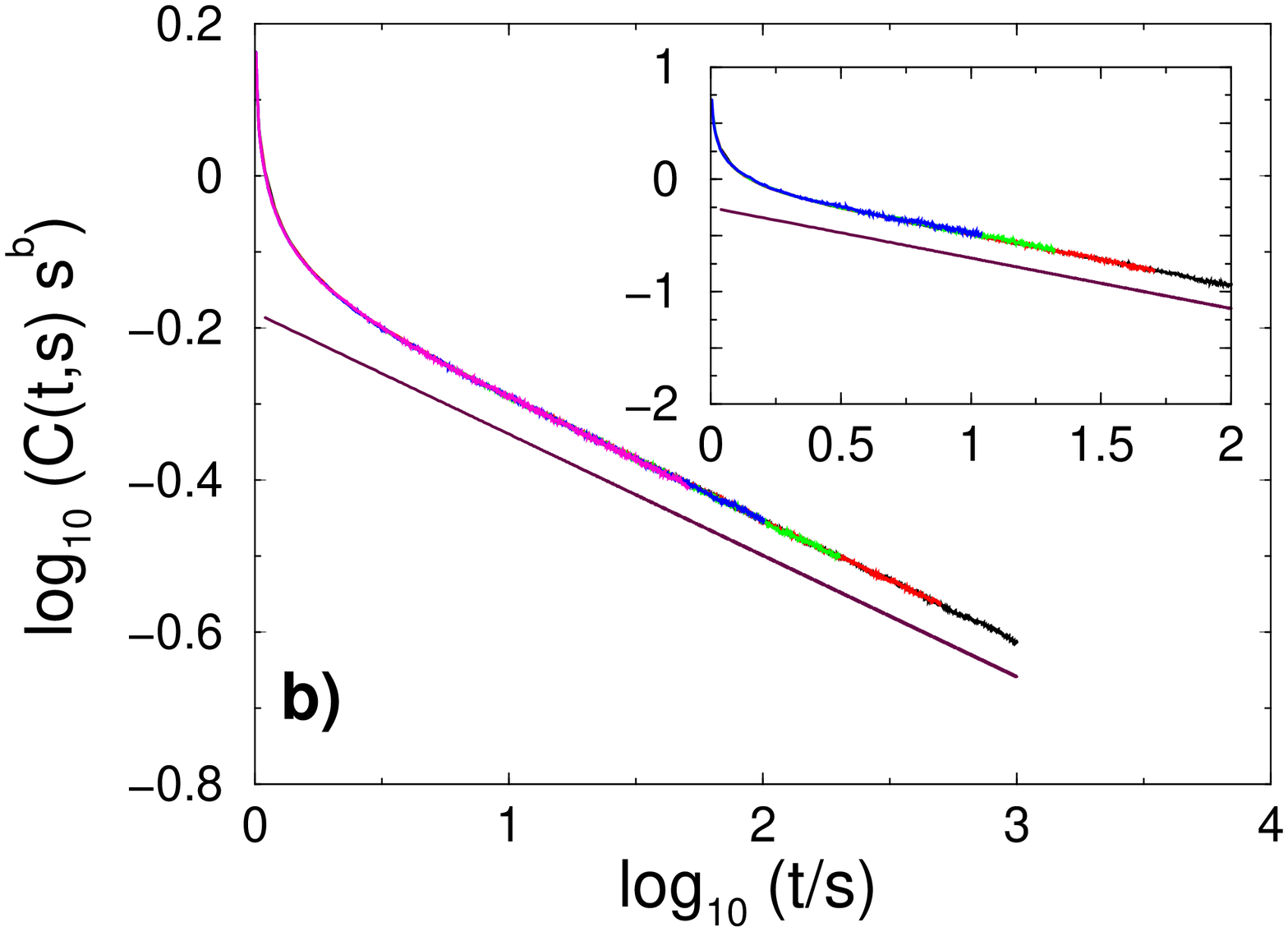}
\caption{a) Two--time autocorrelation function $C(t,s)$ 
for the contact process in $1D$ (main plots) and in
$2D$ (insets). The waiting times are from top to bottom 
$s = [100,200,500,1000]$ in both cases; for one dimension the 
value $s = 2000$ is also included.
These curves are collapsed in b) employing the scaling ansatz of 
Eq. (\ref{Cscal}b,\ref{gl:b}). The straight lines have a slope $-\delta$.}
\label{fig4}
\end{center}
\end{figure} 

The evolution of $C(t,s)$ is shown in Figure~\ref{fig4} for a range of waiting 
times. The collapse of these curves with the scaling 
ansatz of Eq.~(\ref{Cscal}) is also shown. From this collapse, we make
the identification
$\lambda_C/z = \delta$ and also confirm the relation 
$b = 2 \delta$. As expected, we have found a different
asymptotic behaviour for $\Gamma(t,s)$ and $C(t,s)$. 
These results in $1D$ and $2D$ can be
compared with the mean-field prediction in the scaling limit $s\to\infty$ 
with $y=t/s$ kept fixed (see the appendix) 
\begin{equation}
C(t,s) \sim \frac{1}{t \ s} = s^{-2} \left( \frac{t}{s} \right)^{-1} .
\end{equation} 
This is in perfect agreement with the scaling ansatz eq.~(\ref{Cscal}b) since
the mean-field value for the exponent $\delta$ is 
$\delta_{\rm MF} = 1$ (see table 1). 

\begin{table}
\begin{center}
\begin{tabular}{clllc} \hline \hline
Exponent           & $d=1$        & $d = 2$      & mean-field & Reference 
                                                                \\ \hline
$\beta$            & $0.27649(4)$ & $0.583(4)$   & $1$        & \cite{Hinr00}\\
$\nu_\parallel$    & $1.73383(3)$ & $1.295(6)$   & $1$        & \cite{Hinr00}\\
$z$                & $1.58074(4)$ & $1.766(2)$   & $2$        & \cite{Hinr00}\\
$\delta$           & $0.15947(3)$ & $0.4505(10)$ & $1$        & \cite{Hinr00}\\
$\eta$             & $1.50416(7)$ & $1.59(1)$    & $2$        & \cite{Hinr00}\\
$\theta_g$         & $1.50(2)$    & $2.5(5)$     & --         & 
         \cite{Maju96,Hinr98,Muno01}\\ \hline
$b$                & $0.31894(6)$ & $0.901(2)$   & $2$             & \\
$\lambda_\Gamma/z$ & $1.9(1)$     & $2.8(3)$     & --              & \\
$a$                & \hspace{-3.3mm}$-0.57(10)$  
                                  & $0.3(1)$     & $\frac{d}{2}-1$ & \\
$\lambda_R/z$      & $1.9(1)$     & $2.75(10)$   & $\frac{d}{2}+2$ & \\ 
\hline\hline
\end{tabular}
\label{tabI}
\caption{Exponents for the directed percolation universality class in one and 
two dimensions and in mean-field theory. The exponent $b=2\delta$ and 
furthermore $\lambda_C/z=\delta$. The numbers in brackets give the estimated
uncertainty in the last digit(s).} 
\end{center}
\end{table}

\section{The two-time autoresponse function}

\begin{figure}
\begin{center}
\epsfysize=50mm
\epsffile{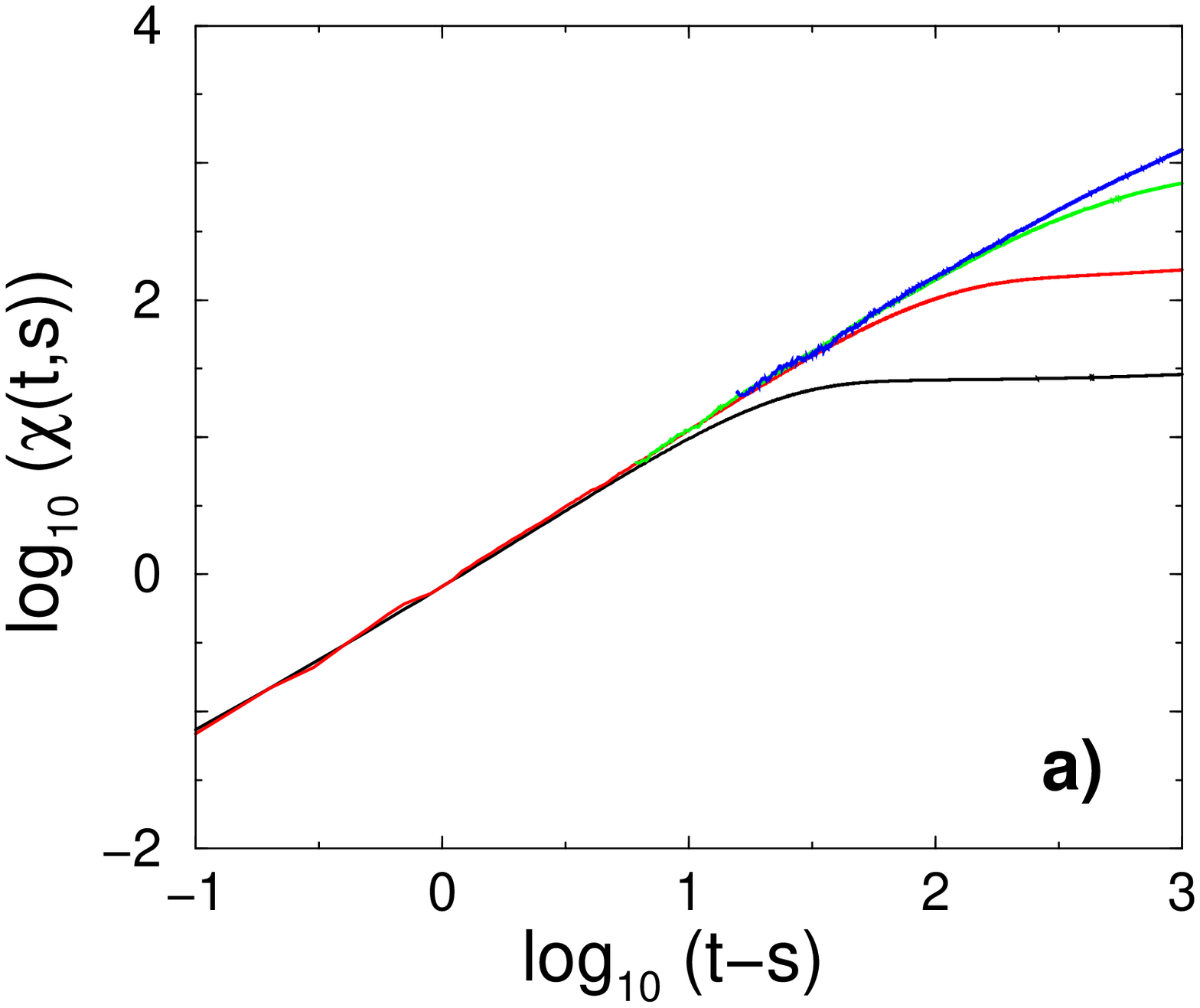}
\epsfysize=50mm
\epsffile{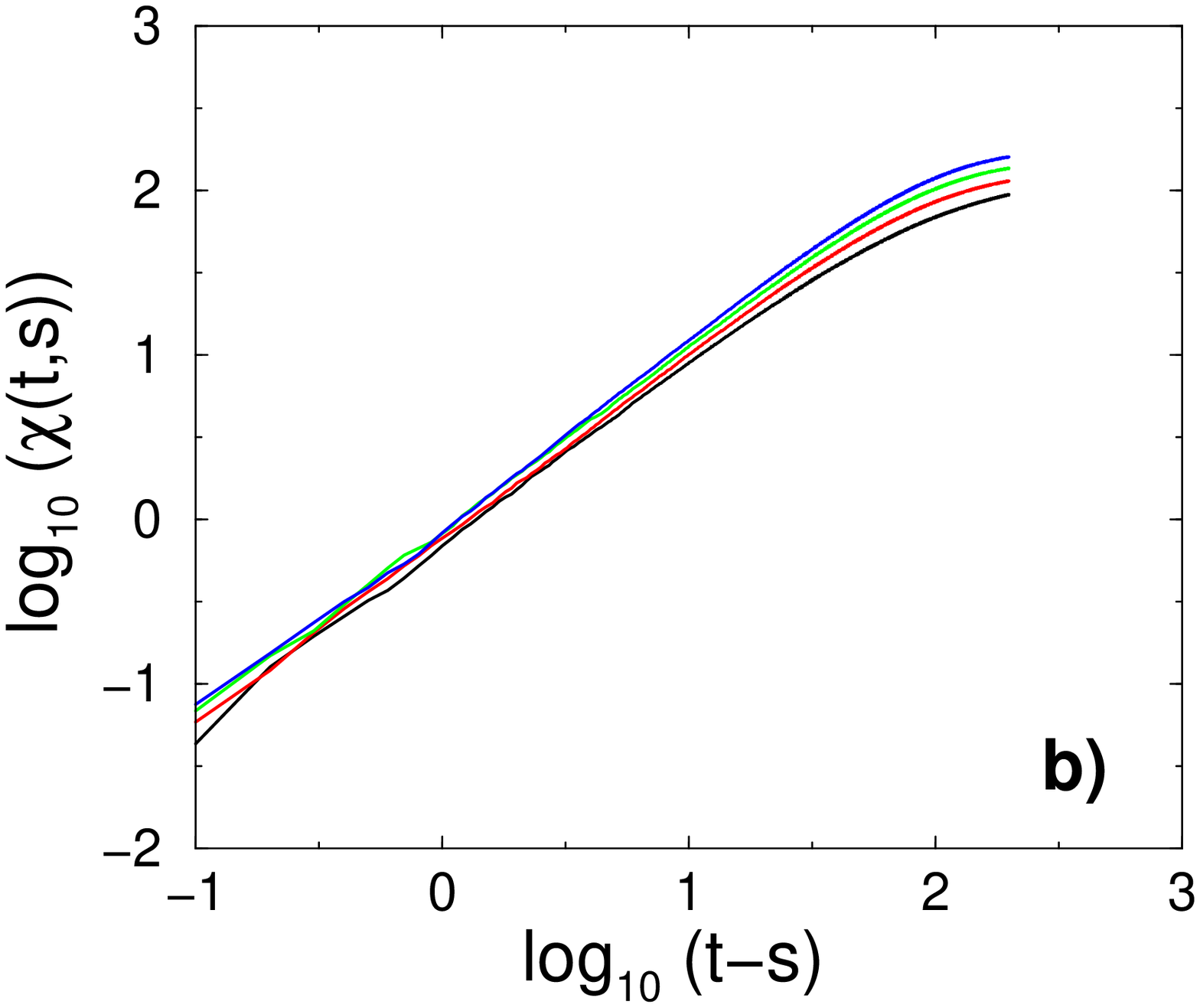}
\caption{a) The ZFC susceptibility $\chi(t,s)$ 
as a function of $t-s$ for the one-dimensional contact process at $p=p_c$.
The external field is changed for the different
curves, from bottom to top $h = [10^{-2},10^{-3},10^{-4},10^{-5}]$, while the
waiting time is maintained at $s = 1000$. In b), the external field is 
fixed at $h = 10^{-3}$ and $\chi(t,s)$ is represented for several 
waiting times, from bottom to top, $s = [250,500,1000,2000]$.}
\label{fig5}
\end{center}
\end{figure}

The linear autoresponse function at time $t$ to an external perturbation by 
the field $h$ at time $s$ is defined as
\begin{equation}
R(t,s) = \left.\frac{\delta\langle n(t)\rangle}{\delta h(s)}\right|_{h=0}.
\end{equation}
Unfortunately, this function is very difficult to estimate directly. 
Hence, as a general rule, the susceptibility 
is measured instead. It is common to work with the zero-field-cooled
(ZFC) susceptibility (we keep the terminology of magnetic spin 
systems by abuse of language)
\begin{equation} \label{4:gl:chi}
\chi_{\rm ZFC}(t,s) = \chi(t,s) = \int_s^t \!\D u\, R(t,u)  .
\end{equation}    
To measure this, one may consider the following dynamical rules. First, 
the 
system evolves according to the original CP rules, until the waiting time
$s$ has elapsed. After that, two copies of the systems, A and B, are kept
evolving in parallel. The copy A continues its evolution without perturbation, 
while copy B is subjected to an external field $h$. This field, 
as explained in section 2,
represents a certain probability of creation ($h > 0$) or destruction ($h <0$)
of a particle whenever the selected site is empty or full, respectively.  
The susceptibility is then calculated as the limit $h \to 0$ of the ratio 
between the difference of the densities of the two copies
and $h$:
\begin{equation}
\chi(t,s) = \lim_{h \to 0} \frac{\langle n_B(t) - n_A(t) \rangle}{|h|} ,
\end{equation}
where $h$ is turned on at time $s$ and kept until time $t$.
However, the application of this method to the contact process is not
straightforward, since a field $h\ne 0$ suppresses the phase-transition and
hence the ageing behaviour. For spin systems, this difficulty may be
side-stepped by using a spatially random magnetic field $h_i=\pm h$
\cite{Barr98}. Although this method may be applied to CP, the random field 
still brings the system out of its critical point, since there is no symmetry 
between the two states of the variable 
$n_i$. Consequently, to measure the susceptibility, there is no 
option but to force the system out of criticality. We have checked  
that applying an uniform field to the whole system is numerically 
more efficient than utilizing a random one. Therefore we shall employ the 
former method in our calculations. We stress that, in both procedures, 
a small enough value of $h$ must be taken to prevent 
saturation for the times considered in the simulations.  
From studies of the magnetization-reversal transition in spin systems
it is known that ageing behaviour can be found if $h$ is 
small enough \cite{Paes03}. In Figure~\ref{fig5}a, the
susceptibility is depicted for a fixed waiting time and several values of the 
external field. The asymptotic behaviour ($h \to 0$), which corresponds to 
the early $\tau = t-s$ regime, may be there clearly observed. In addition, 
in Fig.~\ref{fig5}b, the same
magnitude is represented for a range of waiting times. This shows that there 
is a time window available for studies of the ageing behaviour of the linear
response. 

The interpretation of the zero-field-cooled susceptibility in terms of the
scaling behaviour of the linear autoresponse function $R(t,s)$ meets with
a further difficulty. Na\"{\i}vely, one would simply insert the assumed 
scaling form eq.~(\ref{1:gl:CRskal}) into (\ref{4:gl:chi}). This
simplistic procedure would lead to
\BEQ \label{4:gl:CLZ}
\chi(t,s) \stackrel{?}{=} \int_{s}^{t} \!\D u\, u^{-1-a} f_R(t/u) 
= t^{-a} \int_{1}^{t/s} \!\D w\, w^{a-1} f_R(w)  = s^{-a} f_M(t/s)
\EEQ
However, as pointed out in \cite{Henk03e}, in doing so one neglects the
important condition $t-s\gg t_{\rm micro}$ necessary for the validity of
eq.~(\ref{1:gl:CRskal}). Indeed, it can be shown that taking this condition
into account rather leads to \cite{Henk03e}
\BEQ \label{4:gl:aA}
\chi(t,s) = \chi_0 + s^{-A} g_M(t/s) + \mbox{\rm O}\left(s^{-a}\right)
\EEQ
such that the exponent $A\geq 0$ is in general {\em unrelated} to $a$ and 
where $\chi_0$ is some constant. For example, for phase-ordering spin
systems which are at a temperature above their roughening temperature $T_R$,
one always has $A<a$ \cite{Henk03e}. Hence the term of order 
${\rm O}(s^{-a})$ in (\ref{4:gl:aA}), coming from the integral (\ref{4:gl:CLZ}) 
over the autoresponse function, merely furnishes a finite-time correction
(see the appendix for a related difficulty of using $\chi_{\rm ZFC}$ which
arises in mean-field theory).  

These difficulties may be circumvented by using 
the integrated response in a setup
similar to a {\em Thermoremanent Magnetization} (TRM) experiment, with a
small modification, close in spirit to the `intermediate' protocol
proposed in \cite{Henk03e}. To do so, 
we start with a system evolving with the usual
contact process rules. After a time $s-\tau_c$, the system is split in two
copies, A and B. A continues its evolution without perturbation, whereas B 
is subjected to an external field $h$. The field is 
switched off at time $s$, and from
then on we track the difference between the two copies, which is called the 
thermoremanent density:
\begin{equation} \label{4:gl:rho}
\rho(t,s) = \int_{s-\tau_c}^s \!\D u\, R(t,u) = \lim_{h \to 0}
\frac{\langle n_B(t)-n_A(t) \rangle}{|h|} ,
\end{equation} 
where $h$ is now turned on at time $s-\tau_c$ and switched off at time $s$.
In addition, $\tau_c$ must satisfy the relation $\tau_c \le s$ -- actually, 
this scheme would correspond to a standard TRM experiment if $\tau_c=s$.
In addition, $\tau_c$ must also be much smaller than the relaxation time 
of copy B to avoid that $\rho(t,s)$ becomes a
function of the difference of times $t-s$ alone \cite{Plei03}. 
The scaling of $\rho$ may be obtained from inserting  
eq.~(\ref{1:gl:lsi}) of local scale-invariance into (\ref{4:gl:rho}). It can
be shown \cite{Henk03e} that the leading corrections to scaling are of order 
${\rm O}(s^{-\lambda_R/z})$ which is negligible at criticality. 
 
A last point has to be discussed before the scaling of the integrated response
$\rho$ as calculated here can be analysed quantitatively. Consider
the linear space-time response function
\BEQ
R = R(t,s;\vec{r}-\vec{r}') := \left. 
\frac{\delta\langle n(t,\vec{r})\rangle}{\delta h(s,\vec{r}')}\right|_{h=0}
\EEQ
In contrast to the situation usually considered in magnetism, where the
field $h$ is modelled by spatially uncorrelated random variables, we use
here an uniform magnetic field with spatial correlation, as explained 
before. This implies that
the measured TRM becomes for $s\gg \tau_c$
\BEQ
\rho(t,s) = \frac{1}{L^{2 \, d}}\sum_{\vec{r}}\sum_{\vec{r}'} 
R(t,s;\vec{r}-\vec{r}') = \wht{R}_{\vec{0}}(t,s) ,
\EEQ
where the sums runs over the entire lattice of linear size $L$ and
$\wht{R}_{\vec{k}}(t,s)$ is the Fourier transform of $R(t,s;\vec{r})$
at momentum $\vec{k}$. In order to understand the scaling behaviour of
$\wht{R}_{\vec{0}}(t,s)$, we recall that local scale-invariance predicts for
any given value of $z$ the following form of the space-time response 
\cite{Henk02}
\BEQ \label{4:gl:Rr}
R(t,s;\vec{r}) = R(t,s) \Phi\left( \abs{\vec{r}}/\xi\right) ,
\EEQ
where $R(t,s)$ is the autoresponse function, 
$\xi\sim (t-s)^{1/z}$ and $\Phi$ is a scaling function which can be
obtained from a linear differential equation of fractional order.\footnote{For
$z=2$, it can be shown that $\Phi(u)=\exp(-{\cal M}u^2)$, where $\cal M$ is
a dimensionful constant \cite{Henk02}.} Therefore
\BEQ
\wht{R}_{\vec{0}}(t,s) = \int_{\mathbb{R}^d}\!\D\vec{r}\: R(t,s;\vec{r}) 
\sim R(t,s) \xi^{d} \sim R(t,s) (t-s)^{d/z} ,
\EEQ
where we assumed that the scaling function $\Phi(u)$ falls off sufficiently
rapidly for $u$ large such that the Fourier transform exists. 
If the autoresponse function scales asymptotically (i.e. for $t/s\gg 1$) as
\BEQ
R(t,s) \sim s^{-1-a} \left( \frac{t}{s}\right)^{-\lambda_R/z}
\EEQ
we expect for the TRM the scaling behaviour
\BEQ \label{4:gl:rho0}
\rho(t,s) \sim \hat{R}_{\vec{0}}(t,s) \sim s^{-1-a+d/z} 
\left( \frac{t}{s}\right)^{(d-\lambda_R)/z}
\EEQ

\begin{figure}
\begin{center}
\epsfysize=50mm
\epsffile{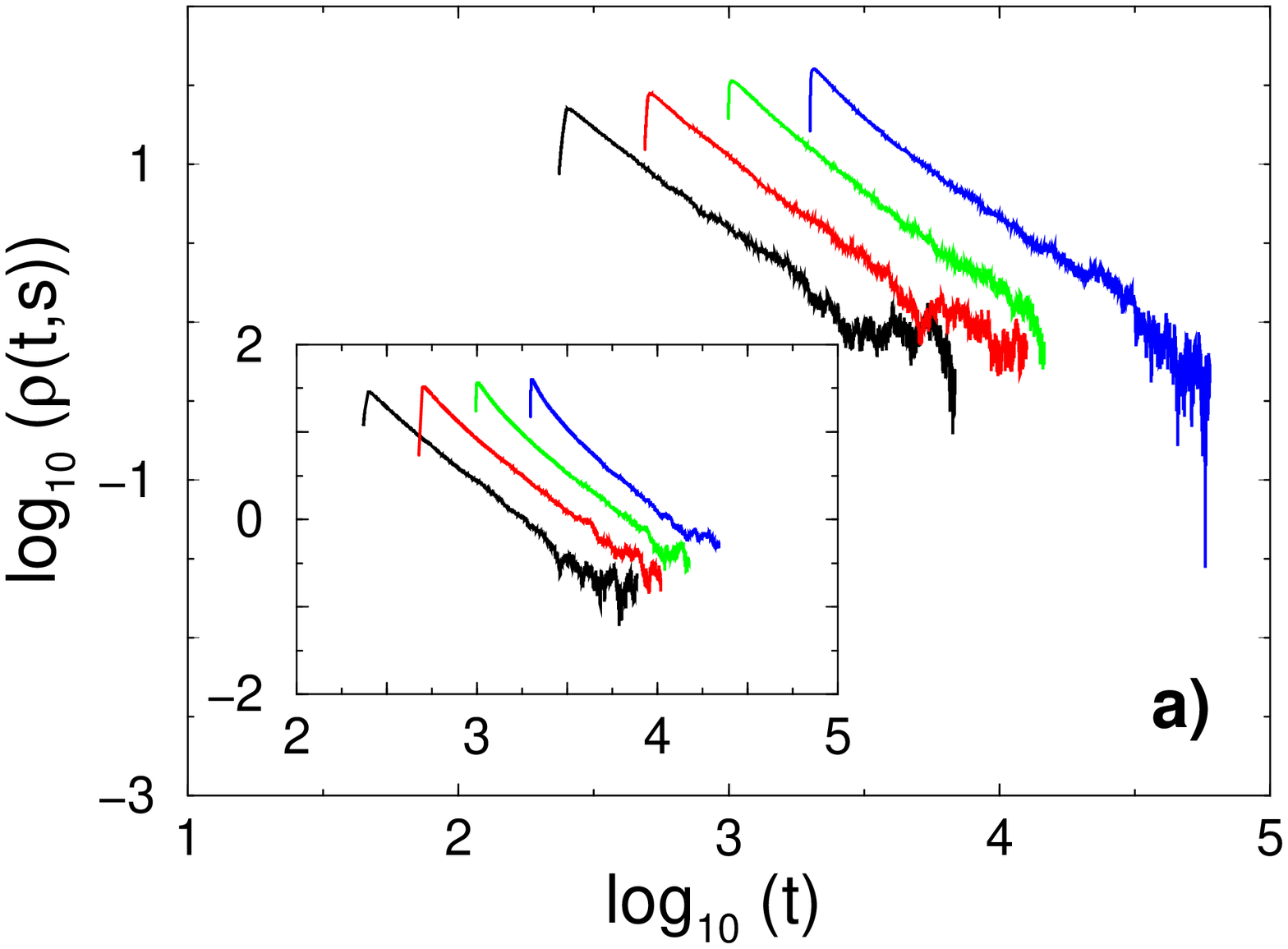}
\epsfysize=50mm
\epsffile{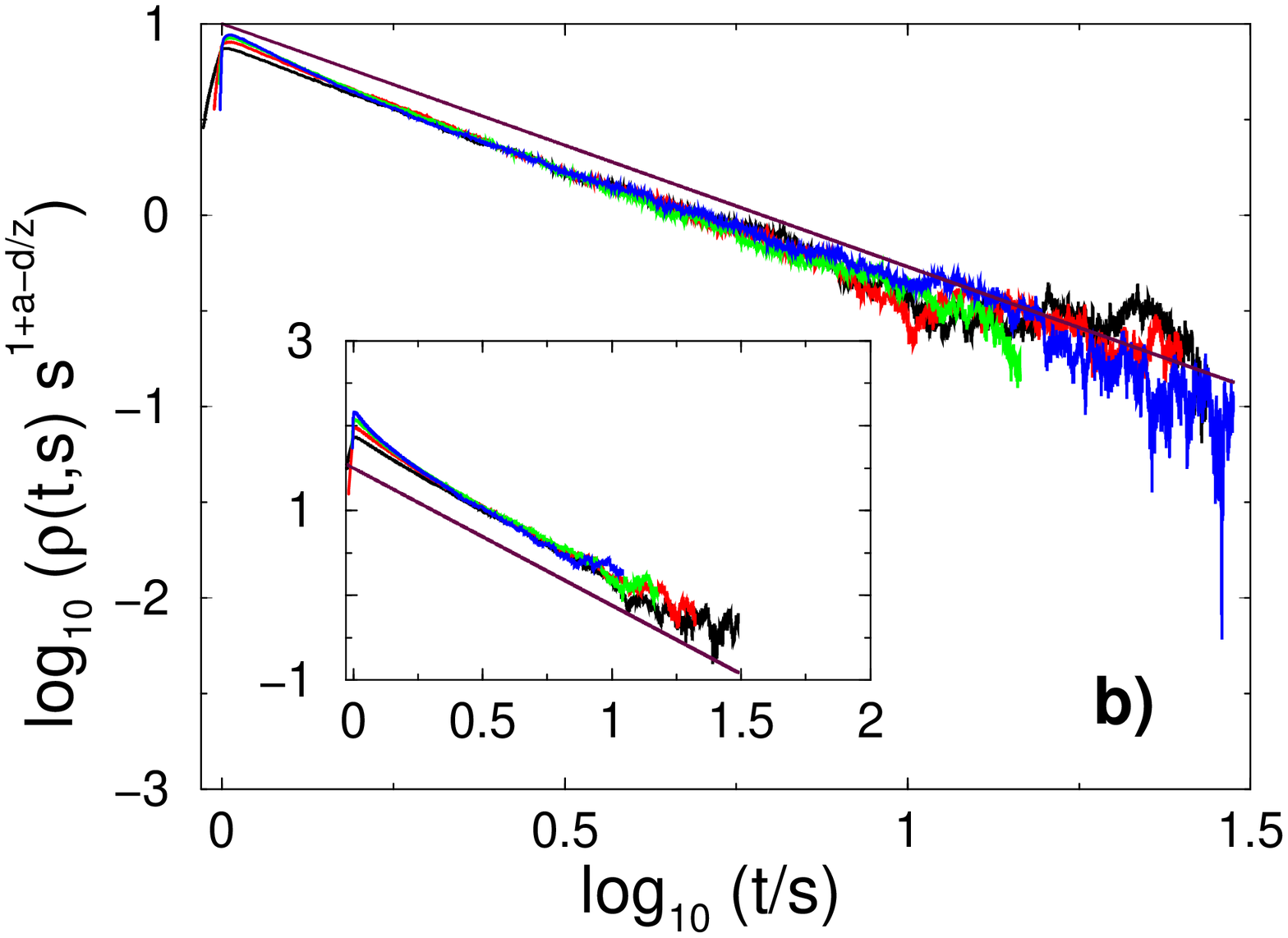}
\caption{a) Temporal evolution of the 'thermoremanent' density 
$\rho(t,s)$ (see \protect{eq.~(\ref{4:gl:rho})}) 
for several waiting times, from bottom to top 
$s=[250,500,1000,2000]$. The main plots are for one dimensional
systems and the insets for $2D$. In b), we collapse the curves of a) 
according to dynamical scaling eq.~(\ref{4:gl:rho0}). 
The slopes of the straight lines are $-1.27$ in $1D$ and $-1.62$ in $2D$.
\label{fig6}}
\end{center}
\end{figure} 

In Figure~\ref{fig6}a, the evolution of $\rho$ is plotted for several 
values of the waiting time. The strength of the field used in the simulation 
is $h = 10^{-3}$ and $\tau_c = 25$. The ageing behaviour is clearly
visible. Replotting the data according to the dynamical scaling behaviour
expected from eq.~(\ref{4:gl:rho0}) (figure~\ref{fig6}b), we find a  nice 
data collapse for the exponent values
\BEQ
1+a-\frac{d}{z} = \left\{ \begin{array}{ll} 
~\hspace{-3.3mm} -0.20(10) & \mbox{\rm ~; in $1D$}\\
~~ 0.17(10) & \mbox{\rm ~; in $2D$} \end{array} \right.
\EEQ
Taking the slopes of the curves in figure~\ref{fig6}b we get
\BEQ
\frac{\lambda_R-d}{z} = \left\{ \begin{array}{ll} 
1.27(10) & \mbox{\rm ~; in $1D$} \\ 
1.62(10) & \mbox{\rm ~; in $2D$} \end{array} \right.
\EEQ
from which the exponent values of $a$ and $\lambda_R$ given in 
table $1$ are obtained. 

It is satisfying that for $d=1$, the results of a complementary 
DMRG calculation \cite{Enss04}
are consistent with ours. Our results may also be compared 
with the prediction of mean-field theory, as derived in the appendix
\begin{equation}
\label{4:gl:Rmf}
R(t,s) \sim \left(\frac{t}{s}\right)^{-2} \left( t-s\right)^{-d/2}.
\end{equation}
The mean-field values for the exponents are then $a = \frac{d}{2}-1$ and 
$\lambda_R/z = \frac{d}{2}+2$. They should become exact for $d>4$ 
and are not too far from the simulational results in lower dimensions.

Comparing with the exponents obtained from the autocorrelation function  in 
section~3, we observe that, within numerical errors, the autocorrelation 
and autoresponse exponents agree
\BEQ
\lambda_{\Gamma} = \lambda_R ,
\EEQ
but that the exponents $a$ and $b$ are different. In fact, within error bars 
they obey the following relation
\BEQ
1+a = b = 2\delta .
\EEQ
This last result leads to important consequences as we shall see in the 
next section. Finally, the uncertainty in the ageing exponents derived from our
numerical data is too large for a direct check of 
the validity of local scale-invariance to be carried out. 
Nevertheless, our results can be 
related to the ones using a direct estimation of $R(t,s)$ via the DMRG 
\cite{Enss04} through the ansatz (\ref{4:gl:Rr}), which is based on local 
scale-invariance. The agreement between the two measurements provides indirect 
evidence for the existence of this symmetry in the CP. On the other hand, the
prediction (\ref{1:gl:lsi}) of local scale-invariance is satisfied by
our mean-field result (\ref{4:gl:Rmf}) and in $1D$, the validity of 
eq.~(\ref{1:gl:lsi}) was directly confirmed from a DMRG calculation
\cite{Enss04}.  

\section{Fluctuation-dissipation theorem}

\begin{figure}[b]
\begin{center}
\epsfysize=50mm
\epsffile{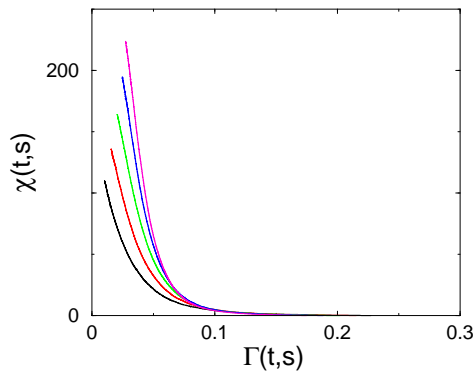}
\caption{The ZFC susceptibility against the connected correlation function 
for a one-dimensional system and with different waiting times, from bottom to 
top $s=[250,500,1000,2000,5000]$. The external field was fixed at 
$h = 7\times 10^{-4}$.} \label{fig7}
\end{center}
\end{figure}

Our observation that $a\ne b$ at criticality has important consequences. 
In particular, it follows that no analogue of the fluctuation-dissipation ratio
can be defined in the critical contact process in a meaningful way. To see this,
we use the scaling ansatz (\ref{Gscal}a) for the connected correlation function
and find
\begin{equation}
\frac{\partial \Gamma(t,s)}{\partial s} = - s^{-b-1} \, 
\left[b \, f_\Gamma(t/s) + (t/s)\, f_\Gamma'(t/s)\right] .
\end{equation}
where the prime denotes the derivative. This implies for a formally defined
fluctuation-dissipation ratio $R(t,s) (\partial\Gamma(t,s)/\partial s)^{-1}$:
\begin{enumerate}
\item a limit fluctuation-dissipation ratio, in analogy with (\ref{Xinfty}), 
can only have a non-trivial value if simultaneously $a=b$ and 
$\lambda_R=\lambda_{\Gamma}$. From table $1$, we see that this is
not the case for the critical contact process. 
\item in contrast to the finding of Sastre et {\it al.} \cite{Sast03} for
the critical voter model, there is no non-trivial limit $t-s\to 0$ 
inside the scaling regime of the fluctuation-dissipation ratio. Existence of
such a limit requires $a=b$ as a necessary condition. From table $1$,
that condition is {\em not} satisfied. Consequently, the supposedly generic
definition of a non-equilibrium temperature --eq.~(\ref{gl:Tdyn})-- as 
proposed in \cite{Sast03} does not extend to the critical contact process and
hence should reflect a peculiarity of the models with 
symmetric states.\footnote{The basic of assumption of \cite{Sast03}, namely 
that $X(t,s)\to 1$ in the short-time limit has been critically reexamined by
Mayer and Sollich \cite{Soll04}, who in the $1D$ Glauber-Ising model at
$T=0$ construct a defect-pair observable such 
that $\lim_{s\to\infty} X(s,s)=3/4$.}  
\end{enumerate}
 
A more graphical way to arrive at the same conclusion is to plot the
ZFC susceptibility $\chi(t,s)$ versus the connected correlation function
$\Gamma(t,s)$: if there were a generalized FDT, we should find a curve with
a constant slope that corresponds to the inverse of the effective 
temperature. We do
so in figure~\ref{fig7} for the $1D$ case, which makes it clear that no
such extension of the FDT exists, in contrast with results \cite{Sast03} 
of the $2D$ critical voter model.

\section{Conclusions}

We have explored the r\^ole of the detailed balance condition in the 
slow non-equilibrium relaxation of a statistical system towards a steady-state. 
In order to do so, we have replaced a key property of physical ageing, namely 
that the basic microscopic processes are reversible, by a property encountered 
in chemical or biological ageing with their underlying irreversible 
microscopic processes. Consequently, the steady-state of the system can no
longer be at equilibrium. One of our main questions was how the accepted 
scaling behaviour found in physical ageing might be modified in the absence
of detailed balance and we tried to get some insight into this through a 
case study of the contact process, which is a paradigmatic example of 
systems with a steady-state phase-transition in the directed percolation
universality class. We have studied the ageing of this model by analyzing
two-time autocorrelation and response functions, obtained from Monte Carlo 
simulations and also analytically from the mean-field approximation. 
Our numerical results are in full agreement with those coming from an 
application of the density-matrix renormalization group to the 
same model \cite{Enss04}. 

Our results reemphasize the importance of having either at least two physically
distinct and stable steady-states or else being at a continuous phase-transition
in order to be able to observe the slow non-exponential relaxation towards
a global steady-state which is associated with the phenomenological aspects
of ageing behaviour, notably the breaking of time-translation invariance and
of dynamical scaling. In this sense, it is natural that in the contact 
process ageing is only found if the steady-state is critical. 
We have shown that the phenomenological scaling description, previously 
developed for ageing magnetic systems, generalizes to the case at hand. 
In particular, we have presented evidence for the exponent relations
\BEQ \label{7:gl:lab}
\lambda_{\Gamma} = \lambda_R \;\; , \;\;
1+a = b = 2\delta
\EEQ
in one and two dimensions and in mean-field theory. 
While the first of those is generic in ageing magnets with an uncorrelated
initial state, the second one is not and suggests that {\em the non-equilibrium 
exponents $a$ and $b$ are in general distinct}.\footnote{For systems with
detailed balance, the steady-state is an equilibrium state where the
fluctuation-dissipation theorem holds. This in turn implies $a=b$ 
\cite{Enss04}.} It will be interesting to see 
whether or not the relations (\ref{7:gl:lab}) hold at the critical points of 
other nonequilibrium universality classes. Eq.~(\ref{7:gl:lab}) implies that a 
recent attempt \cite{Sast03} to define a non-equilibrium temperature through an
extension of the fluctuation-dissipation theorem cannot be generalized beyond 
the rather restricted context where this definition had been proposed. 

\ack

We are grateful to T. Enss, M.A. Mu\~noz, A. Picone and U. Schollw\"ock 
for useful comments. 
This work was supported by the French-Portuguese integrated action under 
project F-$14/04$, and 
from the Portuguese Research Council (FCT) under grants SFRH/BPD/5557/2001 and  
SFRH/BPD/14394/2003.

\appsection{Mean-field theory}

The contact process is described at the continuum level by the Reggeon Field 
Theory (RFT) \cite{Card80}. Let $\phi(t,\vec{r})$ represent the continuum
order-parameter at the spatial
position $\vec{r}$ and at time $t$.  Taking into account the initial 
conditions, which is fundamental to analyze this problem, the RFT-action reads
\BEQ \label{action}
S = \int \!\D t \D\vec{r}\, 
\left[ \tilde{\phi}\left(\partial_t\phi - m\phi -
D\vec{\nabla}^2 \phi\right) + g\phi^2\tilde{\phi} -g \phi{\tilde{\phi}}^2 - n_0
\delta(t) \tilde{\phi} \right]  , 
\EEQ
where $\tilde{\phi}$ is an auxiliary field, $m$ the mass term, 
$D$ the diffusion constant, $g$ a coupling constant and $n_0$ 
the initial density. The mean-field approximation amounts to treating 
(\ref{action}) as a classical action. This yields 
the following equations of motion
\begin{equation}
\frac{\delta S}{\delta\tilde{\phi} (t)}=\left( \partial_t -m -D\nabla^2
\right ) \phi + g \phi^2 -n_0\delta (t)=0
\label{psieq}
\end{equation}
and
\begin{equation}
\frac{\delta S}{\delta \phi (t)}= \left(-\partial_t - m -D\nabla^2
\right )\tilde{\phi} +2 g\phi\tilde{\phi}=0    .
\label{tpsieq}   
\end{equation}
We can now evaluate the classical density and the classical 
response function. The term classical means here averaged with respect to 
action (\ref{action}) but only with the $\phi^2\tilde{\phi}$ vertex 
(diagrammatically, the term $\it classical$ refers to the absence of loops 
in the diagrams). The classical density is given by the sum of all 
tree-diagrams which terminate with a single propagator. If one evaluates these 
diagrams in momentum space, it follows that the $\delta(t)\tilde{\phi}(t)$ 
only make a contribution with $\vec{k}=0$ and therefore all diagrams at tree 
level have ${\vec{k}}=0$, see \cite{Lee94}. The classical density is found 
by Fourier-transforming eq.~(\ref{psieq}). We find
\begin{equation}
\hat{\phi}(\omega)=-\frac{n_0}{m+\II\omega}+
\frac{g}{m+\II\omega}\int \frac{\D\omega'}{2\pi} 
\hat{\phi}(\omega - \omega')\hat{\phi}(\omega')   ,
\label{Fpsieq}
\end{equation}
where $\hat{\phi}(\omega)$ is the Fourier transform of $\phi(t)$. Note that 
the diffusion term has been 
disregarded since it gives no contribution to the final result, as explained
above. Eq. (\ref{Fpsieq}) can be iterated to obtain a perturbative series. For 
$m=0$, that is, at the 
critical point, the series can be summed yielding the exact solution
\begin{equation}
\phi (t) = \frac{n_0}{1+g\, n_0 t}
\label{density}
\end{equation}
Therefore, the classical two-time correlation function is given by
\begin{equation}
C(t,s) = \phi(t)\phi(s)=\frac{n_0^2}{(1 + g\, n_0 s)(1 + g\, n_0 t)}  .
\label{correlation}
\end{equation}

We can now turn to the computation of the response function. The Fourier 
transform of equation (\ref{tpsieq}) with respect to $\vec{r}$ reads
\begin{equation}
-\partial_t \tilde{\phi}({\vec{k}},t)=m\tilde{\phi}({\vec{k}},t) - 
D\vec{k}^2\tilde{\phi}({\vec{k}},t) -2 g \phi(t)\tilde{\phi}({\vec{k}},t)   ,
\label{Ftpsieq}
\end{equation}
where $\tilde{\phi}({\vec{k}},t)$ is the Fourier transform of 
$\tilde{\phi}(t,\vec{r})$. The above equation has the formal solution 
\begin{equation}
\tilde{\phi}({\vec{k}},t)=\tilde{\phi}({\vec{k}},s)e^{(-m+D\vec{k}^2)(t-s)}
\exp\left[2 g\int_{s}^{t}\!\D t'\, \phi(t')\right]   .
\label{fsolution}
\end{equation}
At criticality, $m=0$ and using equation (\ref{density}) 
the above equation yields (setting $D=1$)
\begin{equation}
\tilde{\phi}({\vec{k}},t)=\tilde{\phi}(\vec{k},s) e^{\vec{k}^2(t-s)}
\left(\frac{1+g n_0 t}{1+g n_0 s}\right)^2. 
\label{response}
\end{equation}
In order to calculate the response function is necessary to take $t > s$ and 
to multiply 
both sides of equation (\ref{response}) by $\phi(-{\vec{k}},t+\epsilon)$, 
where $\epsilon > 0$. The 
fact that there is no inertial term in the action (\ref{action}) 
(that is, there are no second-order time derivatives)  
leads to the following boundary condition
\begin{equation}
\lim_{\epsilon\to 0} \,  
\left\langle \phi(-{\vec{k}}, t+\epsilon)\tilde{\phi}({\vec{k}},t)
\right\rangle = \lim_{\epsilon\to 0} \wht{R}_{\vec{k}}(t+\epsilon,t) = 1
\label{boundary}
\end{equation}
Consequently, the classical response function reads
\BEQ
\left\langle \phi(-{\vec{k}},t)\tilde{\phi}({\vec{k}},s) \right\rangle 
=  \wht{R}_{\vec{k}}(t,s)
=e^{-\vec{k}^2(t-s)}\left(\frac{1+ g n_0 s}{1+ g n_0 t}\right)^2
\label{responsef}
\EEQ
In real space, the space-time-dependent response function becomes
\BEQ
R(t,s;\vec{r}) =\left\langle\phi(t,\vec{r})\tilde{\phi}(s,\vec{0})\right\rangle
= R(t,s) \exp\left( -\frac{\vec{r}^2}{4(t-s)}\right)
\EEQ
where the autoresponse function is
\BEQ
R(t,s) = \left(\frac{\pi}{t-s}\right)^{d/2} 
\left( \frac{1+g n_0 s}{1+gn_0 t}\right)^2
\EEQ

We are interested in these expressions in the ageing regime $s\gg 1$,
$t-s\gg 1$. The the autocorrelation function reads
\BEQ
C(t,s) \sim s^{-1} t^{-1} = s^{-2} \left(\frac{t}{s}\right)^{-1}
\EEQ
and we read off the mean-field values $b_{\rm MF}=2$ and
$(\lambda_C/z)_{\rm MF}=2$ of the exponents (mean-field theory being gaussian,
$z_{\rm MF}=2$ is evident). The autoresponse function becomes
\BEQ \label{A:Rcm}
R(t,s) = \pi^{d/2} (t-s)^{-d/2} \left(\frac{t}{s}\right)^{-2}
\EEQ
in agreement with the form predicted by local scale-invariance and we
have the exponent $a_{\rm MF}=\frac{d}{2}-1$ and 
$(\lambda_R/z)_{\rm MF}=\frac{d}{2}+2$. While simple mean-field does not yield
a prediction for the connected correlation function $\Gamma(t,s)$, at least
it furnishes expressions for the correlator and the linear response and
their exponents which should become exact for $d>4$.

We finally illustrate the importance of taking {\em both} conditions 
$s\gg 1$ and $t-s\gg 1$, or equivalently $s\gg 1$ and $t/s\gg1$, 
for the validity of the scaling form (\ref{1:gl:CRskal}) into account. 
First, for the 
thermoremanent protocol (we take $\tau_c=s$) we find, using (\ref{A:Rcm})
\BEQ
\rho_{\rm TRM}(t,s) = \int_{0}^{s}\!\D u\, R(t,u) \sim
t^{1-d/2} \int_{0}^{s/t} \!\D w\, w^2 (1-w)^{-d/2}
\EEQ
which produces a well-defined scaling function for all values of $d$. 
On the other hand, for the zero-field-cooled protocol we have
\BEQ \label{A:chiZFC}
\chi_{\rm ZFC}(t,s) = \int_{s}^{t}\!\D u\, R(t,u)  \stackrel{?}{\sim}
t^{1-d/2} \int_{s/t}^{1} \!\D w\, w^2 (1-w)^{-d/2}
\EEQ
which diverges for all $d\geq 2$. This means that the scaling form 
(\ref{A:Rcm}) {\em cannot} be used near the upper integration limit
in (\ref{A:chiZFC}) and the contribution of the region around $w=1$
must be evaluated separately, without appealing to dynamical scaling
(we remind the reader that the condition $t-s\gg 1$ is implicit in the
continuum limit used to derive (\ref{responsef})). 

After submission of this work, we became aware of a paper by Oerding and
van Wijland \cite{Oerd98} where the two-time connected autocorrelation
function $\widehat{\Gamma}_{\vec{0}}(t,s)$ of reggeon field-theory 
is worked out in momentum space to one-loop order. They find a scaling behaviour
quite analogous to the one observed here. 


\end{document}